\documentclass[journal]{IEEEtran}

\usepackage{graphicx}
\usepackage{algorithm}
\usepackage{algorithmic}
\usepackage{amsmath}
\usepackage{amssymb}
\ifCLASSINFOpdf
\else
\fi
\begin{document}
%
\title{A Bounded Derivation Method for the Maximum Likelihood Estimation on the Parameters of Weibull Distribution}
%
%
%

\author{DeTao~Mao,  
        and  ~Wenyuan~Li,~\IEEEmembership{~Fellow,~IEEE}
\thanks{Wenyuan~Li is with Grid Operations, BC Hydro, Burnaby, BC V3N 4X8,
Canada. e-mail: wenyuan.li@bctc.com.}
\thanks{DeTao~Mao is with ECE
Department, University of British Columbia, Vancouver, B.C.,
Canada V6T 1Z4  e-mail: detaom@ece.ubc.ca.}
}

%
%

\markboth{Journal of \LaTeX\ Class Files,~Vol.~6, No.~1, January~2007}%
{Shell \MakeLowercase{\textit{et al.}}: Bare Demo of IEEEtran.cls for Journals}
%



\maketitle

\begin{abstract}
For the basic maximum likelihood  estimating function of the two
parameters Weibull distribution, a simple proof on its global
monotonicity is given to ensure  the existence and uniqueness of
its solution. The boundary of the function's first-order
derivative is defined based on its scale-free property. With a
bounded derivative, the possible range of the root of this
function can be determined. A novel root-finding algorithm
employing these established results is proposed accordingly, its
convergence is proved analytically as well. Compared with other
typical algorithms for this problem, the efficiency  of the
proposed algorithm is also demonstrated by numerical experiments.
\end{abstract}


\section{Introduction}\label{sec:introduction}

The Weibull distribution \cite{Weibull1951paper} is an important
distribution in reliability and maintainability analysis?. The
estimation on its parameters has been widely discussed, and there
are several methodological categories for this parameter
estimation issue \cite{Cohen1991} \cite{Meeker1998}
\cite{WyL2005}. For example, the graphic methods
\cite{Mann1974WblPlotting}, transcendental equation-solving method
applying bifurcation algorithm \cite{WyL2005}, maximum likelihood
estimation (MLE) method \cite{Cohen1965} \cite{MLE1965a}
\cite{MLE1965b}. The graphic methods, such as  Weibull probability
plotting (WPP) \cite{GraphicMethod4WPP}, are straight forward, but
can not give a precise estimation. The transcendental
equation-solving method has a closed form, it can avoid the
computing expense in the iterative computation of the raw data,
but in order to solve the transcendental equation, advanced
mathematical techniques are required.

In the MLE-based methods, as the basic estimating equation is not
in closed form, it  can be solved only numerically. There are
several typical MLE-based methods for solving this equation, such
as the secant method, the bisection method and the Newton-Raphson
method. However, in both the secant method and the bisection
method, the convergence rates are very low; in the Newton-Raphson
method, it has to  compute both the basic estimating function and
its derivative \cite{NRmethodbyQiao} at each iterative step. In
some cases, the Newton-Raphson method cannot ensure convergence
\cite{NRcounterExample}. Furthermore, these above MLE-based
methods require either initial values or trial computation of the
estimated parameters.

In \cite{Dobson2006}, the author claims that the existence and
uniqueness of the solution  of the basic estimating equation under
MLE method cannot be assured. However, in
\cite{MLEonWeibullDistribution2008} \cite{FarnumProof1997}, based
on Cauchy-Schwarz inequality, the authors have given similar
proofs of the existence and uniqueness on the solution of the
MLE-based estimator. In this paper, we present  a straight forward
proof by mathematics induction, which is much simpler than both
the proofs given  in
 \cite{MLEonWeibullDistribution2008} \cite{FarnumProof1997}.

For the basic estimating function, we have proved the scale free
property of its first order derivative, the  boundary of this
derivative can thus be defined. Moreover, with a bounded
derivative,   at each  iterative step,  the possible range of the
root of the basic estimating function can be determined. We thus
propose a novel MLE-based root-finding algorithm based on these
properties, its computational efficiency and advantages are well
demonstrated by numerical experiments.

The remainder of this paper is organized as follows. In Section
\ref{sec:global monotonicity}, a proof on the monotonicity of the
basic estimating function is given. In Section \ref{sec:asymptotic
property}, the scale-free property of the first order derivative
of this function is proved, the boundaries of the function itself
and its first order derivative are defined. 
 The feasible range of its root thus can be determined.
 In Section
\ref{sec: Designed Agorithm 4 MLE}, by employing these proved
results, a novel root-finding algorithm is hence designed. The
convergence of the proposed algorithm is proved. In Section
\ref{sec: Designed Agorithm 4 MLE}, the performance of this
proposed root-finding algorithm are demonstrated by numerical
experiments. Conclusions are given in
 \ref{sec: conclusion}.
\section{The Global Monotonicity of the Basic Estimating Function}\label{sec:global monotonicity}
\subsection{The Basic Estimating Function for the Two Parameter Weibull Distribution}
\hspace*{\parindent}The density function of the two parameter
Weibull distribution is:
\begin{equation}
f(x)=(\frac{k}{x})(\frac{x}{\lambda})^k e^{-(\frac{x}{\lambda})^k}
\; (x \geq 0, k>0, \lambda>0)\label{eqn:weibull Density function}
\end{equation}

For a sampled data of  $n$ observations with the above
Eqn.\ref{eqn:weibull Density function} as the applicable density
function, its likelihood function is
\begin{equation}
L(x_1,\cdots,x_n;k,\lambda)=\prod_{i=1}^n
(\frac{k}{x_i})(\frac{x_i}{\lambda})^k
e^{-(\frac{x_i}{\lambda})^k}
\end{equation}\label{eqn:weibull likelihood function}
By MLE method, the following equations can be obtained:
\begin{eqnarray}
\frac{\partial \ln L}{\partial k}=\frac{n}{k}+\sum_{i=1}^n \ln x_i
-\frac{1}{\lambda^k}\sum_{i=1}^n x_i^k \ln x_i =0,\\
\frac{\partial \ln L}{\partial
\lambda}=\frac{k}{\lambda}(-n+\frac{1}{\lambda^{k}} \sum_{i=1}^n x_i^k )=0.\label{Eqns: for MLE}
\end{eqnarray}
by eliminating $\lambda$, we get
\begin{equation}
\frac{\sum_{i=1}^n x_i^k \ln x_i}{\sum_{i=1}^n x_i^k }
-\frac{1}{k}=\frac{1}{n}\sum_{i=1}^n  \ln x_i \label{eqns: MLE
basic equation}
\end{equation}
by the above Eqn.\ref{eqns: MLE basic equation}, we can get the
value of $k$ by related numerical algorithms. With $k$ determined,
$\lambda$ can be calculated by Eqn.\ref{Eqns: for MLE} as
\begin{equation}
\hat{\lambda} =  (\frac{\sum_{i=1}^n
x_i^{\hat{k}}}{n})^{\frac{1}{\hat{k}}}= \sqrt[\hat{k}]
{\frac{1}{n}\sum_{i=1}^n x_i^{\hat{k}}}\label{eqn:for lambda}
\end{equation}
we  can therefore calculate both $k$ and $\lambda$. Here $\hat{k}$
$(\hat{\lambda})$ refers to  maximum likelihood estimators for
parameter $k$ ($\lambda$).

 By Eqn.\ref{eqns:
MLE basic equation}, we can define the basic estimating function
$F(k)$ as
\begin{equation}
F(k)=\frac{\sum_{i=1}^n x_i^k \ln x_i}{\sum_{i=1}^n x_i^k }
-\frac{1}{n}\sum_{i=1}^n  \ln x_i-\frac{1}{k}\label{eqn: basic
estimating equation}
\end{equation}

Thus  $F(k)=0$ here is defined as the basic estimating equation.
 By mathematics induction, with
$n\geq 1$ and $k>0$, a simple proof on the global monotonicity of
$F(k)$, i.e., $\frac{\partial F(k)}{\partial k} > 0$ can be given
in the following section. 
\subsection{Proof on the global monotonicity of
$F(k)$}\label{subsec:proofonBasic_Estimating_Equation} Since
\begin{eqnarray}
\frac{\partial F(k)}{\partial k}=\frac{1}{k^2}+ \hspace{4.5cm} \nonumber \\
\biggl\{\sum_{i=1}^n x_i^k {\ln}^2 x_i \sum_{i=1}^n
x_i^k-(\sum_{i=1}^n x_i^k \ln x_i)^2\biggl\}({\sum_{i=1}^n}
x_i^k)^{-2}\label{eqn:1stOrdermonotonicity}
 \end{eqnarray}
 to prove $\frac{\partial
F(k)}{\partial k} > 0$, we need only to prove that for $k>0$ and
any $x_i \in R^{+}$,
\begin{equation}
P_1(x_i,n,k)=\sum_{i=1}^n x_i^k {\ln}^2 x_i \sum_{i=1}^n x_i^k-
(\sum_{i=1}^n x_i^k \ln x_i )^2 \geq 0 \nonumber
 \end{equation}
for $n=1$
\begin{equation}
P_1(x_i,1,k)= x_1^k {\ln}^2 x_1\cdot x_1^k - x_1^{2k}\cdot {\ln}^2
x_1 =0
\end{equation}
for $n=2$
\begin{equation}
P_1(x_i,2,k)= x_1^k x_2^k ({\ln} x_2-{\ln x_1})^2 \geq 0
\end{equation}
for $n=m-1$ ($m\geq 3$, $m \in N^+$), suppose
\begin{eqnarray}
P_1(x_i,m-1,k) \geq 0
\end{eqnarray}
then for $n=m$
\begin{eqnarray}
P_1(x_i,m,k)=P_1(x_i,m-1,k)+ \hspace{2.50cm}  \nonumber \\
 x^k_{_{m}}\sum_{i=1}^{m-1} x_i^k(\ln x_{_{m}}-\ln x_i)^2 \geq 0
\end{eqnarray}
Therefor $\frac{\partial F(k)}{\partial k}>0$, and function $F(k)$
is global monotonic. With the global monotonicity of $F(k)$, the
existence and uniqueness of the root of $F(k)=0$  can be assured.

\section{Boundaries of the Basic Estimating
Function and its First-order derivative}\label{sec:asymptotic
property}
\subsection{Boundaries of the Basic Estimating Function
$F(k)$}\label{subsec: boundary of F(k)}
\begin{figure}
\centering \includegraphics[width=2.750in,height=2.0in, bb=70pt
410pt 400pt 640pt] {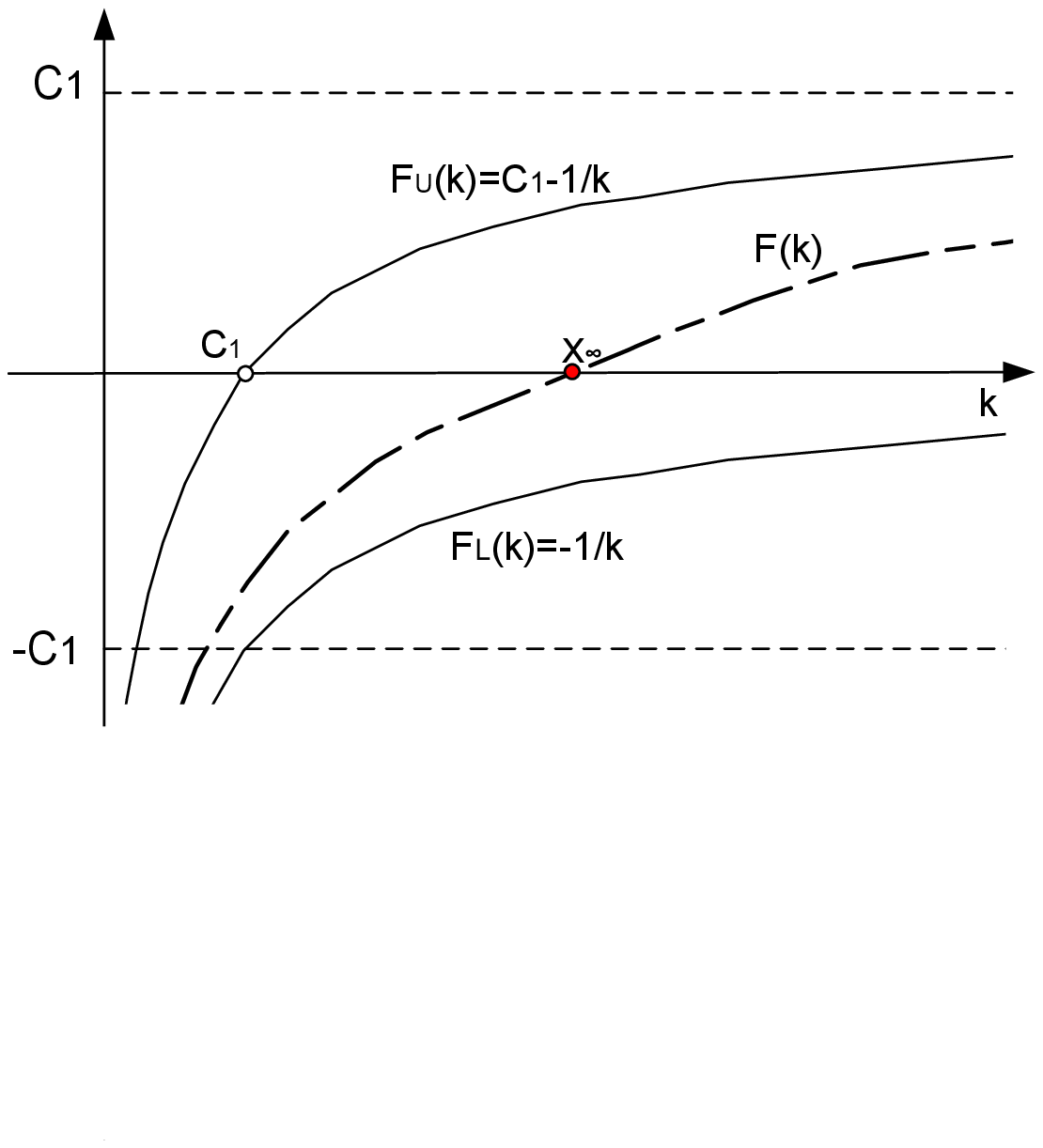} \caption{The boundaries of
$F(k)$: $ -\frac{1}{k}<F(k)<-\frac{1}{k}  + C_1$. Here
$C_1=\frac{1}{n}\sum_{i=1}^{n-1}\ln \frac{ x_{max} }{ x_i}$,
$k>0$. } \label{fig:bounded FK}
\end{figure}

 Let\begin{equation}  G(k) =\frac{\sum_{i=1}^n x_i^k
\ln x_i}{\sum_{i=1}^n x_i^k } -\frac{1}{n}\sum_{i=1}^n  \ln x_i \;
\end{equation}
as
\begin{displaymath}
  \left\{ \begin{array}{l}
G(0)=0 \\ 
\\
G(+\infty)=\frac{1}{n}\sum_{i=1}^{n-1}\ln \frac{ x_{max} }{ x_i}>0
\end{array} \right.
\end{displaymath}
also because
\begin{equation} \frac{\partial G(k)}{\partial k}=P_1(x_i,n,k)(\sum_{i=1}^n x_i^k)^{-2}\geq
0,\hspace{.1cm}
\end{equation}
thus $\forall k>0$
\begin{equation} -\frac{1}{k} \leq  F(k)   \leq  C_1 -\frac{1}{k}
\label{ineqn:rangeof Fk}
\end{equation}
here
\begin{equation}
C_1=G(+\infty)=\frac{1}{n}\sum_{i=1}^{n-1}\ln \frac{ x_{max} }{
x_i}\hspace{.1cm}
\end{equation}

we define the lower boundary of $F(k)$ is curve: ${F_2}(k)=
-\frac{1}{k}$, the upper boundary of $F(k)$ is curve: ${F_1}(k)=
-\frac{1}{k}+C_1$. Then the boundary curves of $F(k)$ can be seen
in Fig.\ref{fig:bounded FK}.


\subsection{Boundaries of the First Order derivative: $\frac{\partial F(k)}{\partial
k}$}\label{subsec:boundaryof1stoderderivative}
 Let\begin{eqnarray}  H(n,x_i,k) = \biggl\{
\sum_{i=1}^n x_i^k {\ln}^2 x_i \sum_{i=1}^n x_i^k \hspace{2.015cm}
 \nonumber \\ - \sum_{i=1}^n x_i^k \ln x_i\sum_{i=1}^n x_i^k {\ln}
x_i \biggl\}({\sum_{i=1}^n} x_i^k)^{-2}
\end{eqnarray}

\textbf{Theorem A}:  $ \forall \, n\in N^{+}, \forall \, x_i >0$,
and $
\forall \, k>0$, 
$\frac{\partial F(k)}{\partial k} \in [k^{-2}, k^{-2}+\ln^2
(\frac{x_{_{max}}}{x_{_{min}}})]$.

PROOF: For a certain $  \lambda_j
>0$ satisfying $\ln (\lambda_j \cdot x_i) \geq 0 $,
\begin{eqnarray}    H(n,  x_i,k)=H(n,\lambda_j \cdot
x_i,k) \hspace{2.45cm} \nonumber \\
\leq \max \{H(n,\lambda_j \cdot
x_i,k)\} \hspace{1.45cm} \nonumber \\
= \ln^2 (\lambda_j \cdot x_{_{max}}) -\ln^2 (\lambda_j \cdot
{x_{_{min}}}) \hspace{0.20cm} \nonumber  \\ =\ln
\frac{x_{_{max}}}{x_{_{min}}}\,\,\ln (\lambda^2_j \cdot x_{_{max}}
 \cdot{x_{_{min}}}) \hspace{0.525cm}
\end{eqnarray}


Since  to satisfy
\begin{displaymath}
\left\{ \begin{array}{l}
\ln (\lambda_j \cdot x_i) \geq 0 \\
\\
\ln (\lambda^2_j \cdot x_{_{max}}\cdot
 {x_{_{min}}}) \geq 0
\end{array} \right.
\end{displaymath}
 the minimum value of
$\lambda_j$ is $x^{-1}_{_{min}}$, i.e., $\lambda_j \in [x^{-1}_{_{min}},+\infty]$. \\

let $\lambda =\gamma \cdot x^{-1}_{_{min}} (\gamma\geq 1)$,
\begin{equation}   \max_{\substack{ n\in N^{+}, x_i>0,  \\ k>0, \gamma\geq 1 }} \{H(n,\lambda \cdot x_i,k)\} =\ln
(\frac{x_{_{max}}}{x_{_{min}}}) \ln (\gamma^2
\frac{x_{_{max}}}{x_{_{min}}}) \end{equation} therefore the
supermum value of $H(n, x_i,k)$ is
\begin{eqnarray}\sup \{H(n, x_i,k)\}=
\min \{  \max_{\substack{x_i>0, n \in N^{+},\\ k>0,
  \gamma\geq 1 }}
\{H(n,\lambda \cdot x_i,k)\} \} \hspace{0.5cm}\nonumber \\ =
\min_{\substack{\gamma\geq 1}} \biggl \{ \ln
(\frac{x_{_{max}}}{x_{_{min}}})[\ln \gamma^2 + \ln
(\frac{x_{_{max}}}{x_{_{min}}})] \biggl \} \hspace{0.5cm}
\nonumber \\
= \ln^2 (\frac{x_{_{max}}}{x_{_{min}}})   \hspace{4.10cm}\nonumber
\end{eqnarray}
Hence
\begin{equation}H(n,  x_i,k)\leq \ln^2 (\frac{x_{_{max}}}{x_{_{min}}})  \end{equation}
It is proved in Section
\ref{subsec:proofonBasic_Estimating_Equation}, the infimum value
of $H(n,\lambda x_i,k)$ is:
\begin{equation}\inf
\{H(n, x_i,k)\} = 0\end{equation} with $\frac{\partial
F(k)}{\partial k}=H(n,x_i,k) +\frac{1}{k^2}$,
 therefore
\begin{equation}\frac{1}{k^2} \leq   \frac{\partial F(k)}{\partial k}  \leq C_2 +\frac{1}{k^2}
\label{ineqn:rangeof dF_dk}
\end{equation}
here $C_2=\ln^2 (\frac{x_{_{max}}}{x_{_{min}}})$. \hspace{0.5cm}
$\square$

 With the boundary of  $\frac{\partial F(k)}{\partial k}$
defined, the feasible solution range of $F(k)=0$  can be
determined.

\subsection{Possible Range of the Final Solution of Equation $F(k)=0$}\label{subsec: feasible Range} The basic estimating
function $F(k)$ can be written as
\begin{displaymath}
 F(k)=
\int_{k_0}^{k}F^{'}(\tau)d{\tau}
\end{displaymath}
According to  Theorem A,
\begin{displaymath}
 \left\{ \begin{array}{ll}
\int_{k_0}^{k} \tau^{-2} d{\tau}<\int_{k_0}^{k}F^{'}(\tau)d{\tau} < \int_{k_0}^{k}[ C_2+\tau^{-2}] d{\tau} & (\text{$k>k_0$})\\
&\\
\int_{k_0}^{k} \tau^{-2} d{\tau}>\int_{k_0}^{k}F^{'}(\tau)d{\tau}
> \int_{k_0}^{k}[ C_2+\tau^{-2}] d{\tau}  &(\text{$k<k_0$})
\end{array} \right.
\end{displaymath}\\

let denote
\begin{displaymath}
 \left\{ \begin{array}{l}
F_L(k,k_0)= \int_{k_0}^{k} \tau^{-2} d{\tau}\\
\\
F_U(k,k_0)= \int_{k_0}^{k}[ C_2+\tau^{-2}] d{\tau}
\end{array} \right.
\end{displaymath}
thus
\begin{displaymath}
 \left\{ \begin{array}{ll}
F_L(k,k_0)<F(k)<F_U(k,k_0) & \text{$(k>k_0)$}\\
&\\
F_L(k,k_0)>F(k)>F_U(k,k_0) & \text{$(k<k_0)$}
\end{array} \right.
\end{displaymath}
Here $F_U(k,k_0)$ and $F_L(k,k_0)$ can be seen the boundary curves
of $F(k)$ at point $(k_0,F(k_0))$ (see Fig.\ref{fig:boundary
Curves}). With knowing $F(k_0)$, the two boundary curves of $F(k)$
can be determined as,
\begin{displaymath}
 \left\{ \begin{array}{l}
F_L(k,k_0)= k_0^{-1}+F(k_0)-\frac{1}{k}   \\
\\
F_U(k,k_0)= C_2k-\frac{1}{k}+(F(k_0)-C_2k_0+\frac{1}{k_0})
\end{array} \right.
\end{displaymath}

We can see that both $F_U(k,k_0)$ and $F_L(k,k_0)$ are algebraic
functions with simple mathematical forms.

\begin{figure}[!t]
\centering \includegraphics[width=2.750in,height=2.0in, bb=120pt
460pt 330pt 600pt] {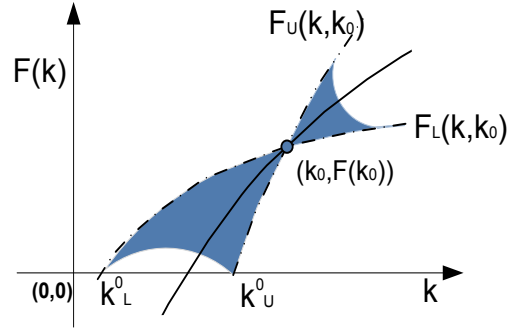} \caption{The two boundary
curves of $F(k)$: $F_L(k,k_0)$ and $F_U(k,k_0)$,  which can
determine  the feasible range of the solution of $F(k)=0$.}
\label{fig:boundary Curves}
\end{figure}

Suppose  $F_U(k,k_0)$ cuts $k$-axe at point $k_U^0$, $F_L(k,k_0)$
cuts $k$-axe at point $k_L^0$, in Fig.\ref{fig:boundary Curves},
it is obvious that the root of equation $F(k,k_0)=0$ must exist in
$[k_L^0,k_U^0]$\, (if $ k_L^0<k_U^0$) or in $[k^0_U,k^0_L]$\, (if
$k^0_L>k^0_U$).

A detailed algorithm  employing the properties in  Section
\ref{subsec: boundary of F(k)},
\ref{subsec:boundaryof1stoderderivative} and \ref{subsec: feasible
Range} will be proposed in the following section, and a proof on
its convergence will be demonstrated as well.

\section{Root-finding Algorithm Design}\label{sec: Designed Agorithm 4 MLE}

To numerically get the solution of $F(k)=0$, an intuitive idea is
to calculate  $F(k1)$ at $k_1=\frac{1}{2}(k^0_L+k^0_U)$, to
further find out a narrower interval $[k_U^{1},k_L^{1}]$ or
$[k_L^{1},k_U^{1}]$. By the same procedure, at
$k_{i+1}=\frac{1}{2}(k^i_L+k^i_U) (i \rightarrow +\infty)$, the
final solution of $F(k)=0$ can be iteratively approximated.

 From the mathematical form of the basic estimating function $F(k)$, we can see that for large $  n, k  $ and
$x_i$, the computational complexity for $F(k)$ is very high, thus
the calculation time of $F(k)$  is an important index in measuring
the efficiency of a root-finding algorithm for $F(k)=0$. A novel
root-finding algorithm has been designed based on these properties
proved in Section \ref{sec:global monotonicity} ,
\ref{sec:asymptotic property} and \ref{subsec: feasible Range}.

\subsection{Convergence of the Bounded derivative Algorithm}
As stated before, a straight forward method employing this idea is
to calculate the final solution iteratively. The feasibility of
this method can be ensured  by the following theorems.

\textbf{Lemma}. For any point $(k_i,F(k_i))$  of function $F(k)$,
the existence and uniqueness of the roots of $F_U(k,k_i)=0$ and
$F_L(k,k_i)=0$ can be assured, therefore the interval
$[k_L^i,k_U^i]\, (i\geq 1)$ that covers the root of $F(k)=0$
always exists.

PROOF: (1) By Inequality \ref{ineqn:rangeof Fk}, we know that
$\forall k>0$, $F(k)> -\frac{1}{k}$. The root of equation $
F_L(k,k_0)= k_0^{-1}+F(k_0)-\frac{1}{k}=0 $ is
\begin{equation}
k_L^0= \frac{1}{F(k_0)-(-k_0^{-1})} >0
\end{equation}
which is reasonable.

(2) Since  $F_U(k,k_0)=0$ can be rewritten as
\begin{equation} C_2k^2+(F(k_0)-C_2k_0+\frac{1}{k_0})k-1=0 \end{equation}
Since here
\begin{displaymath}
 \left\{ \begin{array}{ll}
\Delta =B^2-4AC=(F(k_0)-C_2k_0+\frac{1}{k_0})^2+4C_2 >0 \\

C=-1<0
\end{array} \right.
\end{displaymath}
it is obvious that $F_U(k,k_0)=0$ always has two roots with
different signatures. To satisfy $k>0$, only the positive root
should be preserved.

Therefore, the existence of the interval $[k_L^i,k_U^i]\, (i\geq
1)$ can be assured. \hspace{0.5cm}
$\Box$\\

\begin{figure}
\centering
\rotatebox{0}{\includegraphics[width=3.20in,height=1.90in, bb=85pt
450pt 415pt 630pt] {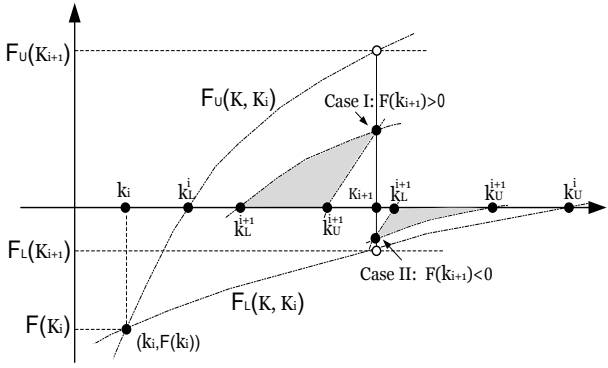}} \caption{Estimation on
the convergence rate of the proposed algorithm with knowing
$F(k_i)$ and  $F(k_{i+1})$, here $F_L(k,k_i)$ and $F_U(k,k_i)$
refer to the boundary functions of $F(k)$ at point $(k_i,F(k_i))$,
$k_{i+1}=\frac{1}{2}(k_L^{i}+k_U^{i})$.} \label{fig: Proof on
Convergence}
\end{figure}

\textbf{Theorem B}. Let  $[k_L^i,k_U^i] $ denote the feasible
interval of the solution of $F(k)=0$ at point $(k_i,F(k_i))$,
$[k_L^{i+1},k_U^{i+1}]$  denote the feasible interval at point
$(k_{i+1},F(k_{i+1}))$, with
$k_{_{i+1}}=\frac{1}{2}(k^{i}_L+k^{i}_U)$, then the convergence
rate
$\gamma=\frac{||k_L^{_{i+1}}-k_U^{_{i+1}}||}{||k_L^i-k_U^i||}<\frac{1}{2}$.
 Here  $k_L^{i+1}=\max(k_L^{i} , k_L^{i+1})$, $k_U^{i+1}=\min(k_U^{i} , k_U^{i+1})$,
 $ i \in N^{+}$.\\

PROOF: With knowing point $(k_i,F(k_i))$ $(i\geq 1)$, we can
deduce  the boundary curves: $F_L(k, k_i)$ and $F_U(k, k_i)$, and
 the related feasible interval $[k_L^{i+1},k_U^{i+1}]$ can  be
determined accordingly (as shown in Fig.\ref{fig: Proof on
Convergence}).

Since if $F(k_{i+1})=0$, then $K_{}$ is the root we are looking
for. Thus here there are only two situations for $F(k_{i+1})$:
$F(k_{i+1})>0$ and $F(k_{i+1})<0$.



\textbf{Case I}: $F(k_{i+1}) >0$

 With knowing $F(k_{i+1})$ $(i\geq 1)$, the boundary
curves $F_L(k, k_{i+1})$ and $F_U(k, k_{i+1})$ at point $(k_{i+1},
F(k_{i+1}))$ can be determined, and so is the related feasible
interval $[k_L^{i+1},k_U^{i+1}]$.

As in this case, by Fig.\ref{fig: Proof on Convergence}, it is
obvious that $k_U^{_{i+1}}<K_{i+1} $, then the convergence rate
\begin{equation}
\gamma = \frac{k_U^{_{i+1}}-k_L^{_{i+1}}}{k_U^i-k_L^i} \leq
\frac{k_U^{_{i+1}}-k_L^{_{i}}}{k_U^i-k_L^i}< \frac{k_
{i+1}-k_L^{_{i}}}{k_U^i-k_L^i}=\frac{1}{2}
\end{equation}

\textbf{Case II}: $F(k_{i+1}) <0$

Similarly,  as shown in Fig.\ref{fig: Proof on Convergence},  in
this case,  it is obvious that $k_L^{_{i+1}}>K_{i+1} $, then the
convergence rate
\begin{equation}
\gamma = \frac{k_U^{_{i+1}}-k_L^{_{i+1}}}{k_U^i-k_L^i} \leq
\frac{k_U^{_{i}}-k_L^{_{i+1}}}{k_U^i-k_L^i}< \frac{k_U^{_{i}}-k_
{i+1} }{k_U^i-k_L^i}=\frac{1}{2}
\end{equation}

Thus the proposed algorithm is convergence, and its convergence
rate $\gamma <\frac{1}{2}$. $\hspace{0.5cm}$ $\square$

\subsection{Improved Algorithm Combining the Secant Method
and the Bounded derivative Method} It is obvious that at large
scale the bounded derivative algorithm can converge rapidly, at
least at the same rate as the bisection method. While in a small
scale, especially in the linearizable neighborhood around the
final solution point of the basic estimating equation, the secant
method method has a better convergence rate. Therefore it is
intuitive to combine both the two methods together. The flow chart
for the combined algorithm can be seen in
Fig.\ref{fig:flowchart4proposedAlgorithm}.
\begin{figure}[!t]
\centering \includegraphics[width=1.250in,height=3.0in, bb=228pt
385pt 340pt 655pt] {BoundedderivativeAlgorithm.eps} \caption{The
flow chart for the combined algorithm. Here $\delta_1$ is  a scale
threshold, $\delta_2$ is the approximating precision or the
computing halt criterion.} \label{fig:flowchart4proposedAlgorithm}
\end{figure}

As shown in the flow chart in
Fig.\ref{fig:flowchart4proposedAlgorithm}, in the combined
algorithm, at a large scale, at each iterative step, we calculate
$F(k_i)$ one time, then with only basic algebraic calculation, we
find  the possible range of the solution, and then narrow down it
to a smaller range at next step.  When the possible range of the
root comes to a small scale, where the linearization part of the
basic estimating function dominates,  the combined algorithm will
switch to the secant method, since it has a better performance in
this case.

Moreover, compared with the secant method, the bisection method
and the Newton-Raphson method,  which requires initial values of
$F(k)$, no initial  value is required in the combined method.

\section{Numerical Examples}\label{sec: numerical examples}
\subsection{General Cases}
With MATLAB, 1000 times of numerical experiments have been taken
to study the performance of this proposed algorithm. In these
simulations, the shape parameter of Weibull distribution $k \sim
U(0,40]$, the scale parameter $\lambda \sim U(0,40]$, the number
of data $N \sim U[2,1000]$, here $U$ refers to
 Uniform Distribution.

Compared with the secant method, the bisection method as well as
the bi-secant method (which has combined the secant and the
bisection method), the computing complexity (here refers to the
calculation time of $F(k)$) of each method under
various approximating precision $\epsilon$ can be seen in the following Table I.\\
\begin{tabular}
 {l|l|l|l|l}\multicolumn{5}{c}{Table I: Averaged calculation times of $F(k)$}
 \\[5pt]\hline
 \multicolumn{1}{c|} {Appro-} &
\multicolumn{1}{|c|}{Secant} & \multicolumn{1}{|c|}{Bisection}
&\multicolumn{1}{|c|} {Bi-secant} &\multicolumn{1}{|c} {Proposed}   \\
 \multicolumn{1}{c|} {precision $\epsilon$} &
\multicolumn{1}{|c}{method} & \multicolumn{1}{|c|}
{method\cite{Cohen1965}} & \multicolumn{1}{|c|} {Method}
&\multicolumn{1}{|c} {method}
\\ \hline \hspace*{\parindent} $10^{-1}$  & \hspace*{\parindent}9.42 &
\hspace*{\parindent}5.21 & \hspace*{\parindent}3.12 &
\hspace*{\parindent}1.68
\\ \hline \hspace*{\parindent} $10^{-2}$  & \hspace*{\parindent}30.22 &
\hspace*{\parindent}6.33 & \hspace*{\parindent}4.29  &
\hspace*{\parindent}2.58
\\ \hline \hspace*{\parindent} $10^{-3}$  & \hspace*{\parindent}88.5 &
\hspace*{\parindent}11.22 & \hspace*{\parindent}9.39  &
\hspace*{\parindent}4.28
\\ \hline \hspace*{\parindent} $10^{-4}$  & \hspace*{\parindent}130.6 &
\hspace*{\parindent}14.19 & \hspace*{\parindent}11.51  &
\hspace*{\parindent}5.05\\
\hline
\end{tabular}
\subsection{A Constructed Case} \hspace*{\parindent}
The advantage of the proposed algorithm can also be verified by a
concrete case, which is generated from a Weibull Distribution. The
sampled data is in Table II, the comparison of the performance of
different methods  can be seen in Table III.\\
\begin{tabular}
 {l|l|l|l|l|l}\multicolumn{6}{c}{Table II:Sampled Data from a Weibull Distribution}
 \\[5pt]
\hline 2.6144 & 4.1834 & 4.3258 & 4.3496 & 4.3740  & 4.4006
\\ \hline 3.2073 & 4.2573 & 4.3273 & 4.3544 & 4.3828 & 4.4051
\\ \hline 3.9800 & 4.2884 &  4.3334 & 4.3646 & 4.3873 & 4.4123
\\ \hline 4.1767 & 4.3150 &  4.3403 &  4.3698 & 4.3959 & 4.4194
\\ \hline 4.4317 & 4.4919 & 4.4448 & 4.5082  & 4.4623 & 4.5439
\\ \hline  4.4756 & 4.5715 &     &   &   &
\\ \hline
\end{tabular}
\\

\begin{tabular}
 {l|l|l|l|l}\multicolumn{5}{c}{Table III: Calculation times of $F(k)$ of various methods}
 \\[5pt]\hline
 \multicolumn{1}{c|} {Appro- } &
\multicolumn{1}{|c|}{Secant} & \multicolumn{1}{|c|}{Bisection}
&\multicolumn{1}{|c|} {Bi-Secant} &\multicolumn{1}{|c} {Proposed}   \\
 \multicolumn{1}{c|} {precision $\epsilon$} &
\multicolumn{1}{|c}{method} & \multicolumn{1}{|c|} {method} &
\multicolumn{1}{|c|} {Method} &\multicolumn{1}{|c} {method}
\\ \hline \hspace*{\parindent} $10^{-1}$  & \hspace*{\parindent}4 &
\hspace*{\parindent}4 & \hspace*{\parindent}3 &
\hspace*{\parindent}1
\\ \hline \hspace*{\parindent} $10^{-2}$  & \hspace*{\parindent}79 &
\hspace*{\parindent}6 & \hspace*{\parindent}3  &
\hspace*{\parindent}1
\\ \hline \hspace*{\parindent} $10^{-3}$  & \hspace*{\parindent}178 &
\hspace*{\parindent}10 & \hspace*{\parindent}3  &
\hspace*{\parindent}2
\\ \hline \hspace*{\parindent} $10^{-4}$  & \hspace*{\parindent}271 &
\hspace*{\parindent}13 & \hspace*{\parindent}21  &
\hspace*{\parindent}3
\\ \hline \hspace*{\parindent} $10^{-6}$  & \hspace*{\parindent}459 &
\hspace*{\parindent}20 & \hspace*{\parindent}22  &
\hspace*{\parindent}4
\\ \hline \hspace*{\parindent} $10^{-10}$  & \hspace*{\parindent}833 &
\hspace*{\parindent}33 & \hspace*{\parindent}23  &
\hspace*{\parindent}5
\\ \hline \hspace*{\parindent} $10^{-14}$  & \hspace*{\parindent}1206 &
\hspace*{\parindent}46 & \hspace*{\parindent}24  &
\hspace*{\parindent}6
\\ \hline
\end{tabular}
\section{Conclusion}\label{sec: conclusion}
In this paper, to assure the uniqueness and existence of the
solution of  the basic estimating function, we have proved its
global monotonicity  in a very simple way. With proving the
scale-free property of this function's first order derivative, the
possible range of its solution at each iterative step can be
determined. Based on these properties, a novel root-finding
algorithm is proposed in this paper. Its efficiency  has been
demonstrated by numerical experiments.

For future work, this method can  be extended  to type I and type
II data censoring cases.
%
\IEEEpeerreviewmaketitle
\bibliographystyle{IEEEtran} 

\bibliography{MLE4Weibull}
\end{document}